# Search for Dark Matter with LHC proton Beam Dump

Ashok Kumar[a], Archana Sharma[b,*]

*[a]Delhi University, Delhi, India*
*[b]CERN, Geneva, Switzerland*

**Abstract**

Dark Matter (DM) comprising particles in the mass range of a few MeV to GeV is waiting to be explored, given the many theoretical models accommodating cosmological abundance. We hereby propose an experiment with the LHC proton beam of 7 TeV striking onto the beam dump target, emitting neutrinos and possibly, Dark Matter candidates. This experiment would also permit to observe signatures involving elastic and inelastic processes involving DM candidates, electrons and strongly interacting particles present in nuclei of the dump target. There will be residual neutrino background present in each of these signatures, hence the proposed experimental detector sub-systems would be such that they would involve as final states, elastically or inelasticity scattered, standard model particles. The bump or the excess in the tail of the kinematic distributions will eventually give us glimpse of presence of new particles which could possibly be Dark Matter candidates. Given the parameters of the LHC machine, the sensitivity of such an experiment is unique in the energy region it is supposed to exploit.

---

[*]*Corresponding author Archana.Sharma@cern.ch*

## 1. Introduction

Dark Matter (DM) comprising particles in the mass range of a few MeV to GeV is needed to be explored, because of the many theoretical models involving extensions to the Standard Model (SM), which can accommodate DM-SM interactions that yield the observed DM cosmological abundance. On the experimental side, such models predict particles that explain the currently discrepancy in the muon anomalous magnetic moment and resolve other anomalies in astrophysical sector, while simultaneously evading cosmological and direct-production constraints. Recent theoretical work has motivated for having light sub-GeV Dark Matter candidates that interact with ordinary matter through mediator particles of different masses. These scenarios constitute a cosmologically and phenomenologically viable chance to look for Dark Matter. Light mass Dark Matter particles are not easy to detect using traditional methods of DM detection, but can be copiously produced and then detected along with neutrinos. This lead us to search for DM in new experimental approach which is highly complementary to other approaches, such as underground direct detection experiments, cosmic and gamma ray satellite and balloon experiments, neutrino telescopes, and high-energy collider experiments [1-3].

This experiment could be performed by placing a detector after the proton beam dump site. The detection of DM particles is facilitated by the proton beam passing through the shielding material, and depositing visible energy inside the detector by scattering off various target particles or if unstable, by decaying inside the detector volume. In the latter case the detector would be complemented by separation of charged particles and time of flight in the decay volume. We need to examine the area present near the dump site and consequently plan the detail of the site readiness. The run would be completely parasitic without affecting normal operations of the LHC program. The proposed utility is based on the DM candidate produced by default interactions taking place inside the dump target. Neutrinos and alike particles cannot be stopped by any of the dump material. At LHC, there are millions of interactions happening in the TeV energy range so we can naturally benefit from to hjave such an unique experimental facility. We are working on full sub-detector components to optimize the maximally achievable sensitivity.

The technique of searching for light sub-GeV Dark Matter with proton beam dump setups has been extensively highlighted in Snowmass 2013,

BDX, LBNE, and Project X white papers.

## 2. Physics Motivation

The experiment aims to detect Dark Matter particles produced in the beam dump of highest energy, and reasonably high intensity, experimental facility at LHC dump. The overwhelming astrophysical and cosmological evidence supports the existence of DM, however its elementary nature remains still to be unknown. The expectation that Dark Matter has some kind of interaction with Standard Model matter is strongly motivated either as an annihilation product of thermal Standard Model matter in the early Universe or as a product of the same unknown processes that generate the baryon asymmetry. There is currently an active program to probe particle DM scattering with direct detection experiments, annihilation with indirect detection telescopes, and production with particle accelerators. However, most of these efforts are designed to search for heavy (Few GeV to 1 TeV) DM candidates and sharply lose sensitivity to lighter (sub-GeV) states whose signals are either too feeble to detect or lie in high-background regions. In fact, Dark Matter candidates are readily motivated in the entire MeV-to-TeV range. The best-motivated models of MeV-to-GeV-mass Dark Matter are those whose interactions with ordinary matter are mediated by new GeV-scale dark force carriers (for example, a gauge boson can mix with the SM photon) [4-12].

The best constraints on multi-GeV Dark Matter interactions are from underground searches for nuclei recoiling off non-relativistic Dark Matter particles in the Galactic halo. However, these searches are insensitive to few-GeV or lighter Dark Matter, whose nuclear scattering transfers invisibly small kinetic energy to a recoiling nucleus. Direct detection using proton-scattering offers an alternative strategy to search for few GeV Dark Matter.

## 2. LHC Dump System

The nominal LHC beam contains an unprecedented stored energy of 350 MJ, contained in 2808 bunches with a beam sigma of the order of 0.3 mm. Point 'IR6' of the LHC is dedicated to the beam dumping system. The functionality of the beam dump system is working on fast-extraction of the beam in a loss-free way from each ring of the collider and transporting it to an external absorber, positioned sufficiently far away to allow for appropriate beam dilution in order not to overheat the absorber material. A loss-free extraction requires a particle-free gap in the circulating beam, during which the field of the extraction kicker magnets can rise to its nominal value. Given the destructive power of the LHC beam, the dump system meets extremely high reliability criteria, hence the probability of very clean DM signatures.

## 3. Detector Sub-systems

Dark Matter in the MeV - GeV range is hitherto relatively unexplored, and has been a topic of considerable discussion. The most striking signal that this experiment would look for comprises events with an order of ~ few to 100s of GeV range of electromagnetic energy deposition. With the detector and the experimental set-up proposed here, this signal will be easily detected over a negligible background. This striking signature can arise in two classes of models: those where DM scatters elastically off atomic electrons in the detector, and in those where the DM can scatter in-elastically in the detector and subsequently de-excite in the active detector material to GeV-scale electron pairs, leading to electromagnetic energy deposition. It will also be possible to detect small signals produced by light DM particles scattering off a nucleon. However, detection thresholds need to be fixed at values as low as possible ( ~ few MeV), where spurious signals from beam-related (neutrinos) and cosmogenic (muons, neutrons and neutrinos) backgrounds limit the measurement sensitivity of the DM-nucleon scattering channel. Nevertheless the sensitivity of the experiment to a broad range of possible DM interactions could provide a tool for systematics and consistency checks.

The experimental setup will utilize a magnet, a tracking station, a triggering calorimeter and a redundant muon system. The target will be placed on the upstream edge of the analyzing magnet and will be few centimeters away from the first layer of the tracker, both of which will be housed in a vacuum chamber. The particle momenta will be measured using tracking station. The calorimeter may be used as the primary trigger for the experiment as well as for electron identification. This will be followed by the muon spectrometer for detecting muons.

Several directional techniques have been proposed for a directional detection of Dark Matter. The key point is to access to the initial direction of the ion recoiling due to the elastic scattering by a WIMP. We can estimate the direction by exploiting information of the recoil track since the initial direction is preserved in the respective detector components. We can simulate observables to quantify the preservation of the initial direction of the recoiling ion in the detector. A gaseous detector, we wish to implement, seems well suited to achieve the measurement of the direction of WIMP-induced nuclear recoils.

## 4. Simulation, Modeling and possible experiment

An investigation of the beam particle composition, the background seen by the experiment and hence optimization of the particle identification methods and beam instrumentation, a detailed simulation model of the upstream proton beam is needed. All the beam elements (magnets, collimators, vacuum elements and beam detectors) have to be precisely modeled, and massive Monte Carlo simulations for production has to be performed, in order to minimize statistical uncertainties. The model will take into consideration effects like muon-to-gamma production, synchrotron radiation energy loss, Coulomb multiple scattering, etc. Since the two lists depend on different parameterizations of several hadronic cross-sections, the predicted production rate of particles can be quite different depending on the momentum range. As a first step, the proton beam hitting the dump target and producing secondaries involving the hadronic processes will be simulated.

The pion, kaon and proton content of the beam has to be normalized to the total charged particle production (with both positive and negative momenta). In the lower momenta range, electrons dominate the spectrum, while for energies lower than a few GeV, the decay length of pions and kaons becomes a significant factor in the population of those particles reaching the experiment.

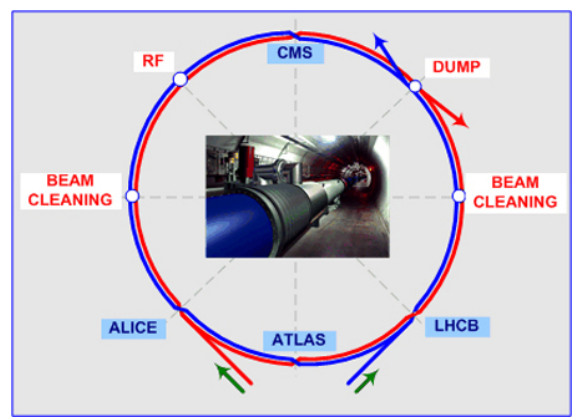

Fig.1. Possible location at P6 for a DM Search Experiment

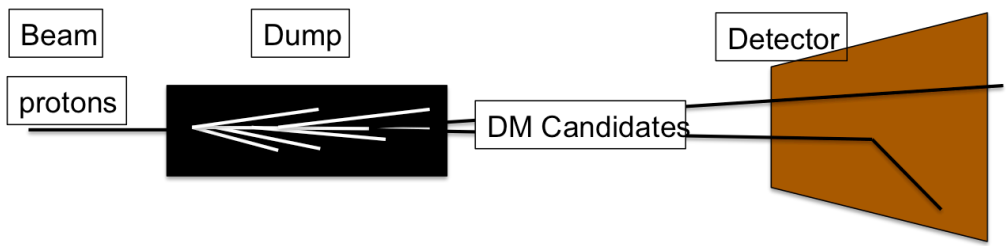

Fig. 1b: Possible set up for Dark Matter search experiment

This DM search experiment could be proposed at one of the proton beam dumps of the LHC, and Point 6 is a possibility, see Fig.1a. A possible experimental set up is shown in Fig. 1b.

## 5. Conclusion

The proton interactions happening at the LHC dump site can be used for Dark Matter detection. The proposed experiment can search for Dark Matter particles in the MeV - GeV mass range probing an entirely new parameter space beyond the reach of existing and proposed experiments. This proposal benefits from the several detector R&D programs ongoing around the globe, especially for LHC Upgrades, hence the experiment can be built with maximum sensitivity exploiting ~ TeV range interactions and cutting edge detector technologies. The site needs to be investigated in coordination with the LHC Accelerator Complex management. After all p-p collisions, neutrinos and Dark Matter particles are produced in millions and pass unnoticed along the beam line, through the dump and into earth. The full experiment could be expected to be ready for installation in the LHC LS3 (2024- 2025) and data taking soon thereafter. A more detailed "Expression of Interest" is in preparation.